\documentclass{elsart1p}
\usepackage{graphics}
\usepackage{graphicx}
\usepackage{epsfig}

\usepackage{amssymb}

\def\gtorder{\mathrel{\raise.3ex\hbox{$>$}\mkern-14mu
 \lower0.6ex\hbox{$\sim$}}}
\def\ltorder{\mathrel{\raise.3ex\hbox{$<$}\mkern-14mu
 \lower0.6ex\hbox{$\sim$}}}
\def\deg{^\circ}

\def\gegm{G_E / G_M}
\def\ge{G_E}
\def\gm{G_M}
\def\gep{G_{Ep}}
\def\gmp{G_{Mp}}
\def\gen{G_{En}}
\def\gmn{G_{Mn}}

\begin{document}

\begin{frontmatter}

\title{Unpolarized nucleon structure studies utilizing polarized
electromagnetic probes}

\author{John Arrington}
\address{Physics Division, Argonne National Lab, IL, USA}

\begin{abstract}

By the mid-1980s, measurements of the nucleon form factors had reached a stage
where only slow, incremental progress was possible using unpolarized electron
scattering.  The development of high quality polarized beams, polarized
targets, and recoil polarimeters led to a renaissance in the experimental
program.  I provide an overview of the changes in the field in the last ten
years, which were driven by the dramatically improved data made possible by a
new family of tools to measure polarization observables.

\end{abstract}

\end{frontmatter}

\section{Introduction}\label{intro}

Elastic electromagnetic form factors provide the most direct insight on
the spatial distribution of the quarks in the proton and neutron.  They encode
information on the distribution of charge and magnetization, and are
thus sensitive to both the distribution and the dynamics of quarks.  Elastic
lepton--nucleon scattering has been the preferred method to
measure the form factors since the 1955 SLAC measurement that provided the
first extraction of the proton radius~\cite{hofstadter55}. Such measurements
were extensively pursued for decades, both on the proton and on the neutron
(using deuterium targets), and by the mid-1980s, this was considered to be a
fairly mature field.  Subsequent measurements in the late 1980s and early
1990s provided mainly incremental improvements to either the precision
or the kinematic coverage of the data.

After four decades of effort, however, there were still large kinematic regions
where only very limited measurements of the form factors were possible. There
were two significant limitations to form factor measurements based on
unpolarized electron scattering: the fact that the cross section is only
sensitive to a specific combination of the form factors and the lack of a free
neutron target.

The form factors represent the difference between scattering from a point-like
object and from a spatially extended target.  For a spin-1/2 target, the cross
section for the exchange of a single virtual photon is sensitive to a
specific combination of the form factors: $(Q^2/4M^2)\gm^2+\varepsilon\ge^2$,
where $\ge$ and $\gm$ are the electric and magnetic form factors, $-Q^2$ is the
four-momentum squared of the virtual photon, $M$ is the mass of the target
nucleon, and $\varepsilon$ is the virtual photon polarization parameter.  The
virtual exchange photon can have a longitudinal contribution in addition to
the transverse contribution of real photons, and so $0<\varepsilon<1$, with
$\varepsilon=0$ corresponding to fully transverse photons (in the limit
$\theta \to 180\deg$), and $\varepsilon=1$ corresponding to the maximal
longitudinal component in the forward scattering limit.  In a ``Rosenbluth
separation'' of the form factors, cross sections are measured at fixed $Q^2$
over a range in $\varepsilon$.  The $\varepsilon \to 0$ limit isolates the
contribution from $\gm$, while the $\varepsilon$ (or $\theta$) dependence
yields sensitivity to $\ge$.  However, the $Q^2$ weighting on the magnetic
form factor makes it difficult to isolate $\gm$ at low $Q^2$, except for
scattering angles approaching 180$\deg$, while $\ge$ is difficult to isolate
at high $Q^2$, where it has only a small contribution to the cross section.

For the neutron, there is an additional limitation, owing to the lack of a
free neutron target.  Early measurements of the neutron form factors came from
inclusive quasielastic scattering from the deuteron, and so were a combination
of the neutron and proton scattering cross sections, thus requiring a large
correction to remove the proton contribution as well as corrections for the
binding and motion of the nucleons.  The proton corrections were eliminated in
later experiments by measuring both the scattered electron and the struck
neutron, but this introduces final state interaction corrections for the
neutron in addition to the nuclear effects for the deuteron target.

\begin{figure}[htb]
\begin{center}
\includegraphics[height=13.0cm,angle=90]{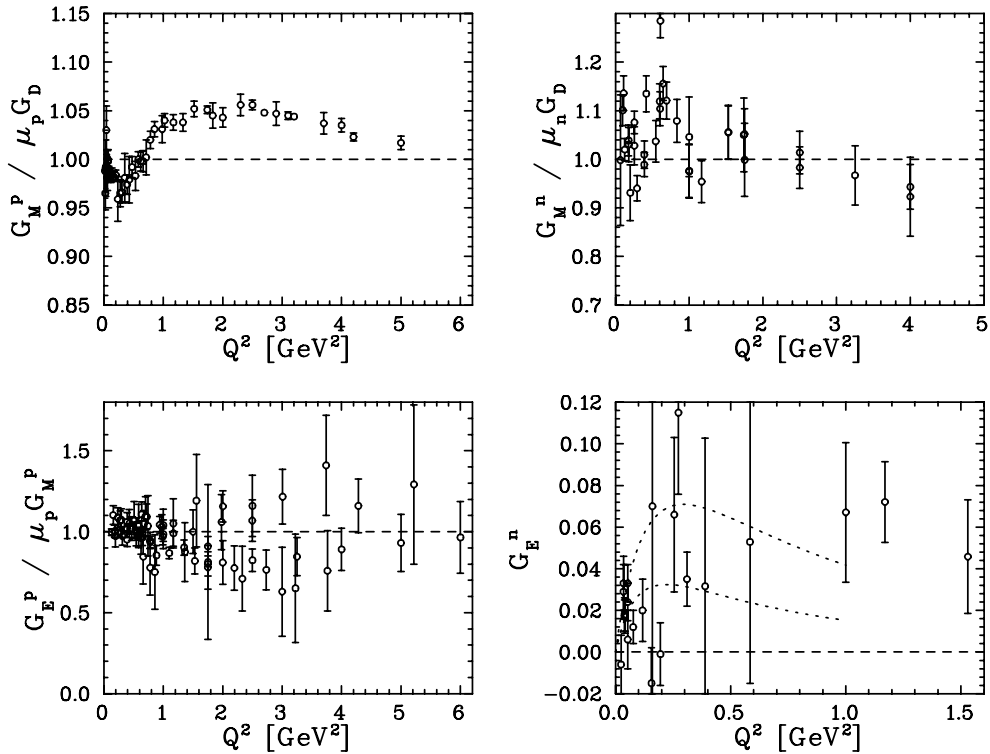}
\caption{Status of form factor extractions as of the late 1990s.  The
proton and neutron magnetic form factors are shown as a ratio to the dipole
fit.}
\label{fig:ffs_old}
\end{center}
\end{figure}

Thus, even in the late 1990s, our knowledge of the form factors was still
rather incomplete.  The proton magnetic form factor, $\gmp$, was well measured
up to $Q^2=30$~GeV$^2$, while the electric form factor, $\gep$, was measured 
to $\sim 6$~GeV$^2$, with much lower precision for $Q^2 > 2$--3~GeV$^2$, as
$\gm$ dominates the cross section at large $Q^2$.  The neutron magnetic form
factor, $\gmn$, was relatively well measured up to $Q^2$ values of
5--6~GeV$^2$, while direct measurements of $\gen$ were mainly upper limits.
Figure~\ref{fig:ffs_old} gives an idea of the status of nucleon form factor
measurements up to 5--6~GeV$^2$ as of the late 1990s.  Measurements of $\gmp$
extend up to 30~GeV$^2$, although with lower precision at higher $Q^2$, while
the other form factors have only very low precision data at higher $Q^2$
values.  For $\gen$, the only data above 1.5~GeV$^2$ was consistent with zero,
with relatively large uncertainties.  The dashed lines indicate the range as
extracted from measurements of e--d elastic scattering, which yields a large
model-dependent uncertainty.  Details on these early form factor extractions
can be found in recent reviews~\cite{gao03,hydewright04,arrington07a,perdrisat07}.

Except for $\gen$, all of the form factors were in approximate agreement
with a dipole fit: $G_D = 1/(1+Q^2/0.71)^2$, with $Q^2$ in GeV$^2$.  In
the textbook interpretation, this implies that the charge and magnetization
distributions would be well described by an exponential distribution,
corresponding to the Fourier transform of the dipole form.  However, this
picture is only valid in the limit of low $Q^2$, as it neglects
model-dependent relativistic boost corrections. Nonetheless, the data suggest
that the proton charge and magnetization distributions are similar, as is the
neutron magnetization distribution.  The neutron electric form factor is zero
at $Q^2=0$, corresponding to the neutron charge.  The positive value of 
$\gen$ at finite $Q^2$ implies a positive charge at the core of the neutron
with a negative cloud of charge at larger distances.  This is consistent with
the ``pion cloud'' picture, where the neutron fluctuates into a proton
and a negative pion.

\section{Polarization Observables}\label{polarization}

During the last decade, the experimental program to measure form factors was
completely reinvented, as it became possible to use polarization degrees of
freedom to make dramatically improved extractions of the form factors.  This
concept was already well understood~\cite{akhiezer58,dombey69,arnold81}, but
such measurements require high polarization, high intensity electron beams, as
well as high figure-of-merit polarized targets or recoil polarimeters.  The
development of these tools made rapid progress in the 80s and 90s, spurred on
initially by interest in nucleon spin structure.  Once they were widely
available, they were used to study a variety of topics unrelated to the spin
structure studies that motivated much of the work.

In particular, polarization measurements dramatically improved the extraction
of nucleon form factors.  Over large kinematic regions, the Rosenbluth
extractions were primarily sensitive to only one of the form factors, and so
could not be used to separate $\ge$ and $\gm$.  In polarization measurements,
either the polarization transfer to the nucleon or the double spin asymmetry in
scattering from a polarized nucleon, the observables depend only on the ratio
of the electric to magnetic form factor: $\gegm$.  By themselves, polarization
measurements cannot provide absolute measurements of either form factor, when
coupled with cross section measurements, they allow for a precise extraction
of both form factors, even in regions where the cross section is sensitive
only to one of the form factors. In the last ten years, the new measurements
utilizing polarization observables have almost entirely supplanted the previous
measurements, and in some cases, yielded surprising new results.

\section{Neutron Form Factors}

The initial focus was on making improved measurements of the neutron.  
Polarization measurements not only allowed for a much better measurement of 
$\gen$, which is nearly impossible to access in a Rosenbluth separation,
but also yielded reduced corrections for the nuclear effects in quasielastic
scattering from the neutron in deuterium or $^3$He.  Recoil polarization
measurements typically use deuterium targets, while polarized target
measurements have most frequently used $^3$He targets.  While polarized $^3$He
targets have a larger contamination from protons, the two protons spend most
of the time with opposite spins, which means that the polarization of the
neutron is very similar to the polarization of the $^3$He nucleus, and that
the dilution due to scattering from the anti-aligned protons is the largest
correction.

\begin{figure}[htb]
\begin{center}
\includegraphics[height=5.0cm,angle=0]{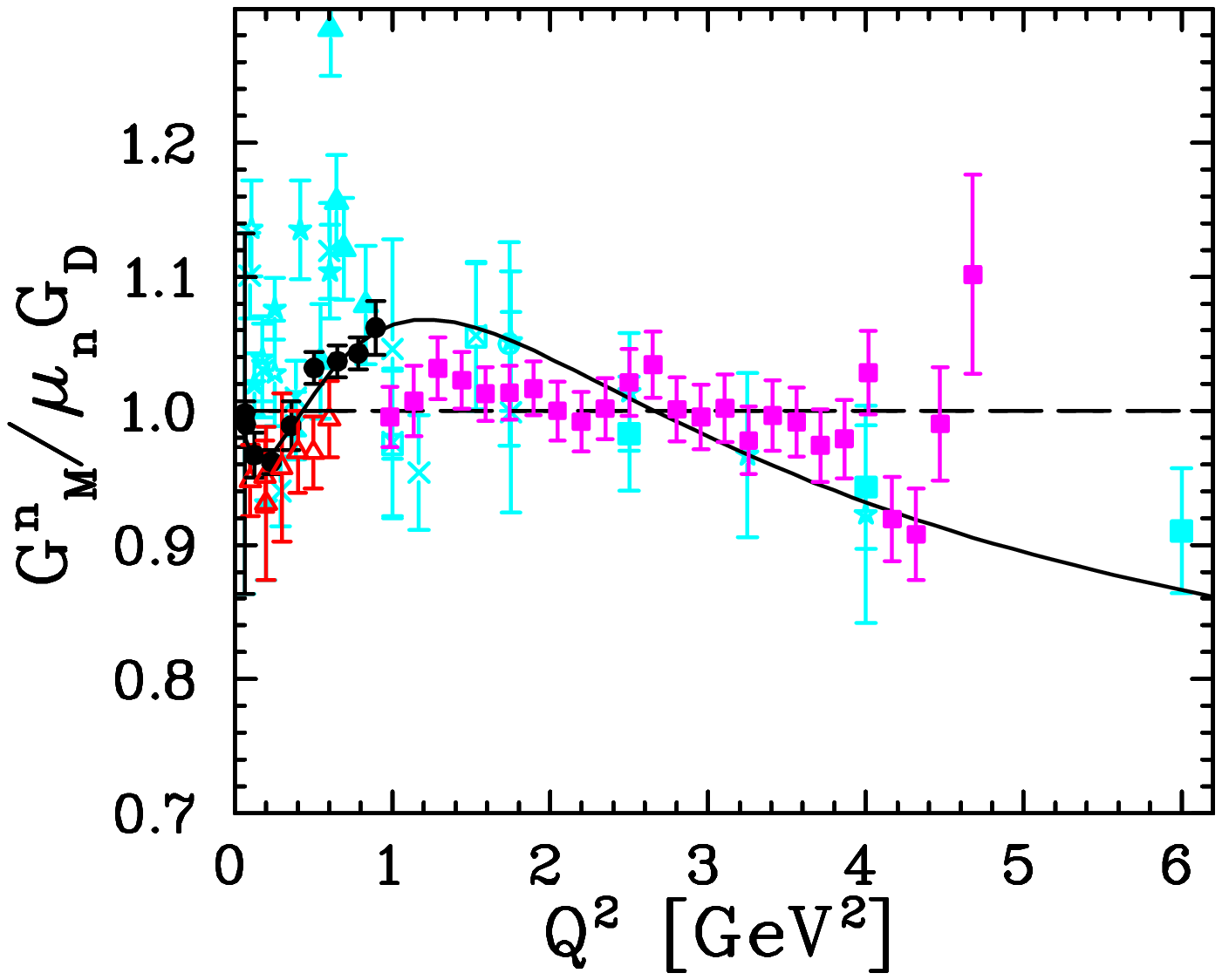}
\includegraphics[height=5.0cm,angle=0]{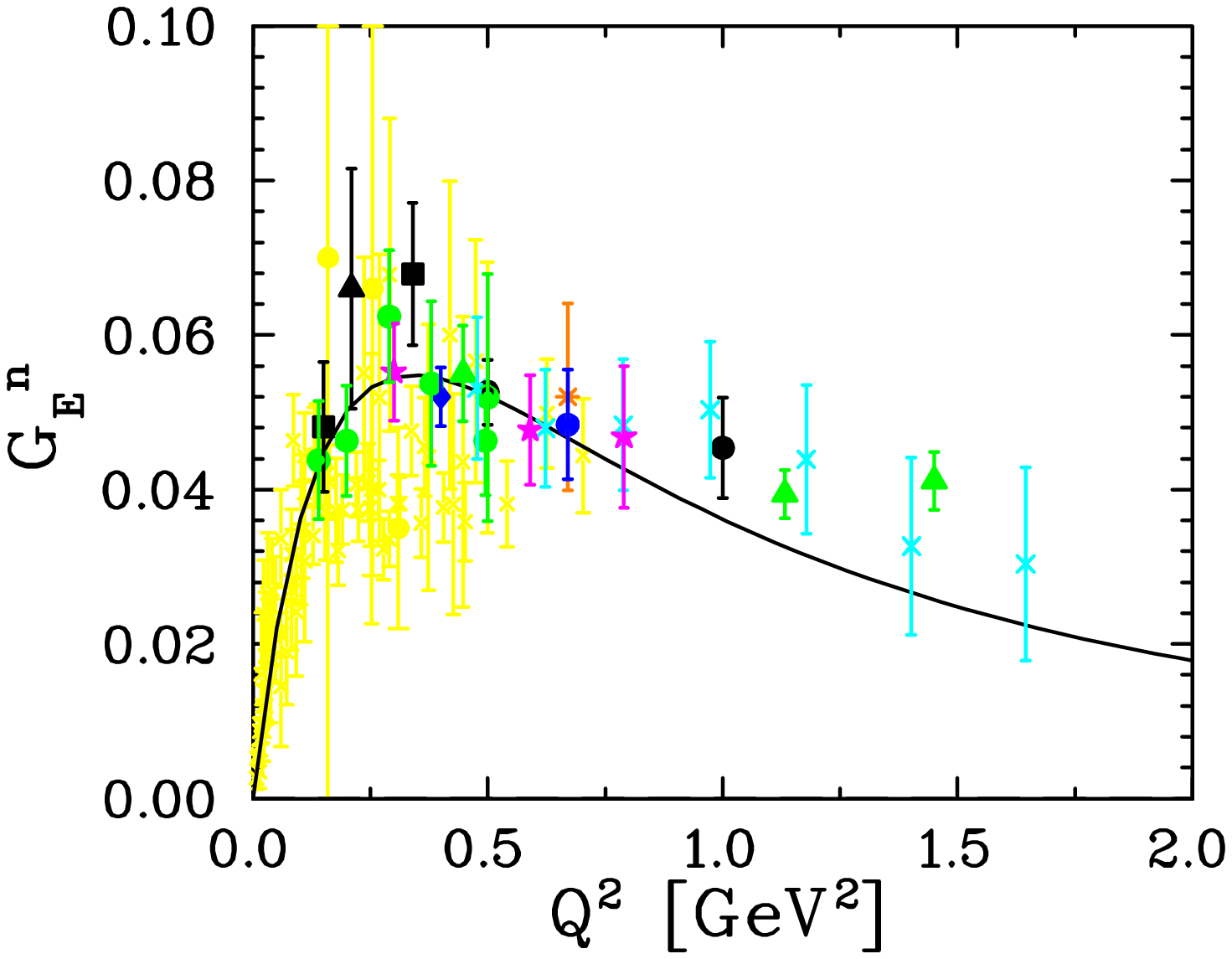}
\caption{Present status of neutron form factor extractions.  For $\gmn$,
the cyan (light grey) points come from inclusive or coincident quasielastic
scattering from deuterium, The hollow red triangles are measurements on
polarized $^3$He~\cite{anderson07}, the solid black circles and solid magenta
squares are from ratio measurements~\cite{anklin98,lachniet08}.  For $\gen$,
the light gray (yellow and cyan) points are from elastic e--d scattering,
while the other points are a variety of polarization measurements.  Above 
2~GeV$^2$, only upper limits on $\gen$ currently exist, but data on polarized
$^3$He have been taken up to 3.4~GeV$^2$ at Jefferson Lab and are currently
under analysis.  Figures adapted from Ref.~\cite{arrington07a}}
\label{fig:neutron}
\end{center}
\end{figure}

Unpolarized measurements of the ratio of proton to neutron knockout were also
used to improve our knowledge of $\gmn$.  Taking the ratio of the
$^2$H(e,e'p) to $^2$H(e,e'n) cross sections removes the large proton
contamination that is limits the inclusive measurements, while
almost entirely cancelling the nuclear effects.  The largest issue is then a
precise measurement of the efficiency of the neutron detector, along with a
correction for the difference between final state interactions of the proton
and neutron.  At large $Q^2$, the ratio measurements have provided much
improved precision on the extraction of $\gmn$, while at low $Q^2$, both
techniques have been used.  The extraction of $\gmn$ is somewhat unusual, as
the polarization observable for scattering from a nucleon are sensitive only
to the ratio $\gegm$, and nuclear targets are required to extract $\gmn$.  For
a free neutron target, the asymmetry depends only on $\gegm$, and by selecting
the angle of the spin relative to the beam polarization, one can choose a term
that is roughly proportional to $\gegm$, or a term that depends mainly on the
kinematics, with only a small correction from $\gegm$.  In the latter case,
the fact that $\gegm$ is small for the neutron means that the asymmetry for
scattering from a neutron is already known.  Thus the resulting asymmetry for
inclusive scattering, where the neutron is not tagged, is simply the neutron
asymmetry diluted by scattering from the (nearly unpolarized) protons in
$^3$He.  This dilution depends only on the relative cross section for
scattering from the neutron and the protons, and thus the asymmetry is
actually used to extract the same cross section ratio that is directly
measured in the ratio technique, but using an entirely different technique.

For $\gen$, a range of different measurements of A(e,e'n) have been performed
to verify the techniques and to test nuclear corrections.  Measurements of
recoil polarization in scattering from deuterium yield results that are 
consistent with measurements of double spin asymmetries made from polarized
deuterium and $^3$He targets. Figure~\ref{fig:neutron} shows the present
status of $\gmn$ and $\gen$ extractions.  The data sets included are described
in more detail in Ref.~\cite{arrington07a};  the results from the CLAS
high-$Q^2$ ratio measurement of $\gmn$ are taken from a recent
preprint~\cite{lachniet08}.

It is clear from Fig.~\ref{fig:neutron} that the polarization and ratio
measurements had a huge impact, dramatically improving the precision for
$\gmn$, and providing essentially all of the high precision extractions of
$\gen$.  For the proton, the situation was expected to be improved, but not as
dramatically, as the main limitation was the extraction of $\gep$ at high
$Q^2$.  However, the initial measurement of $\gep$ at high $Q^2$  didn't just
improve the precision for $\gep$, it also showed that the ratio $\gep/\gmp$
fell with increasing $Q^2$.  This surprising behavior motivated follow-up
experiments meant to verify the results and extend the extraction of $\gep$ to
higher $Q^2$.

\section{Proton Form Factors and Two-Photon Exchange}

While much of the effort went into the neutron measurements, the extraction
of $\gep$ from Rosenbluth extractions was of limited precision at high $Q^2$.
In addition, there were questions about the consistency of the results from
different measurements.  The initial proton measurements were aimed at
clarifying the situation and improving the precision of $\gep$ extractions
above 1--2~GeV$^2$.

\begin{figure}[htb]
\begin{center}
\includegraphics[height=5.0cm,angle=0]{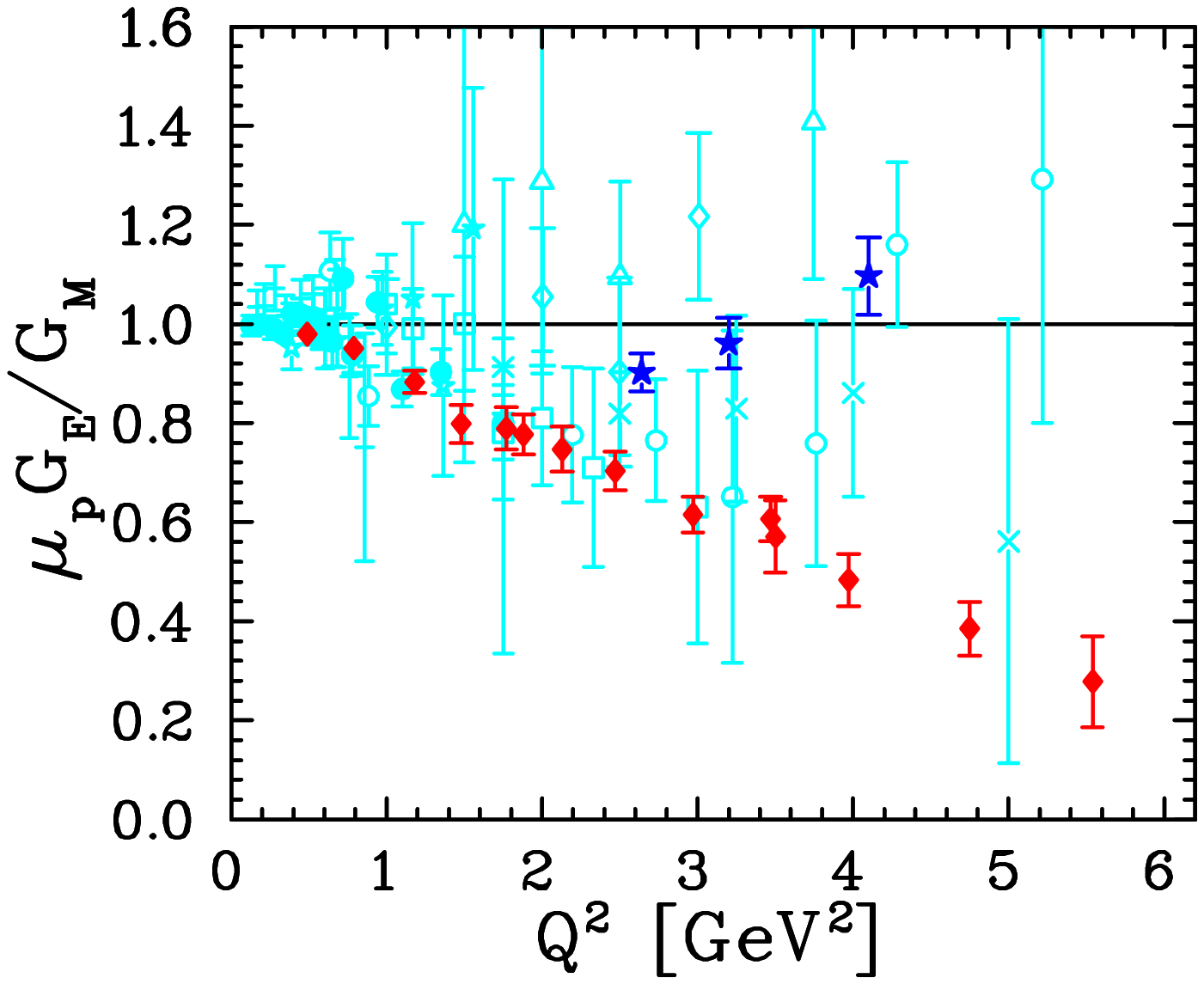}
\includegraphics[height=5.0cm,angle=0]{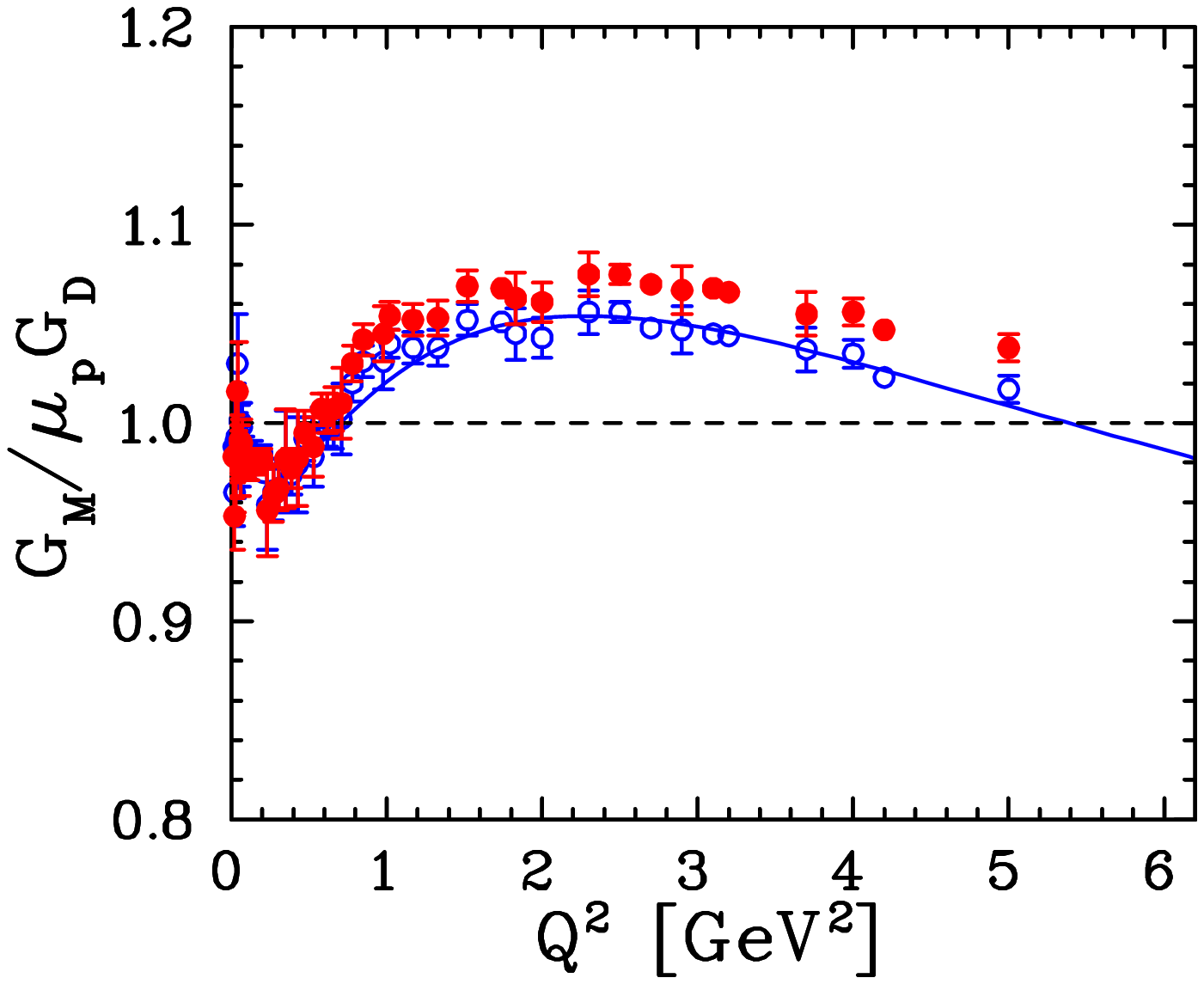}
\caption{Present status of the proton form factor extractions.  The left
plot shows measurements of $\gep$, with the cyan (light gray) symbols
showing results from previous Rosenbluth extractions, the solid red diamonds
showing the high $Q^2$ polarization measurements~\cite{jones00,gayou02}, and
the solid blue stars showing the new ``Super-Rosenbluth''
extractions~\cite{qattan05}.  The right plot shows the extractions of
$\gmp$ from Rosenbluth extractions before (blue open circles) and after
(red filled circles) corrections for two-photon exchange~\cite{arrington07c}.}
\label{fig:proton}
\end{center}
\end{figure}

Figure~\ref{fig:proton} shows the Rosenbluth and polarizations results for
$\gep$.  While the Rosenbluth measurements were of poorer quality, there
appeared to be a real discrepancy with the new polarization measurements.  A
reexamination of the world's cross section data showed that there were no
significant inconsistencies within the set of previous data, and demonstrated
that the discrepancy with polarization was well beyond the uncertainties of
the data sets, even taking into account the normalization and correlated
systematic uncertainties of the cross section data~\cite{arrington03a}.  A new
``Super-Rosenbluth'' measurement, using proton rather than electron detection
to provide extremely small relative systematic uncertainties verified the
discrepancy~\cite{qattan05}.  The experiment yielded uncertainties comparable
to the polarization data, as shown in Fig.~\ref{fig:proton} (solid stars).

Today, it is believed that the discrepancy is due to two-photon exchange
corrections to the Born cross section.  While early calculations had suggested
these corrections were very small, a reexamination of two-photon exchange
(TPE) in light of the new experimental results showed that a relatively small
correction, with a strong angular dependence, could significantly modify the
Rosenbluth extraction of $\gep$ and explain the
discrepancy~\cite{guichon03,blunden03,chen04,arrington04a}.  This led to a
significant effort to examine the impact TPE corrections could have on both
nucleon form factors and a range of other
observables~\cite{arrington04a,blunden05b,arrington07b,carlson07}.  With the
assumption that TPE corrections explain the entire discrepancy, it has been
possible to make updated extractions of the proton form
factors~\cite{arrington07c} using hadronic calculations of the TPE
corrections~\cite{blunden05a}.  The additional uncertainty associated with
these corrections does not appear to be a limiting factor in the form
factor extractions, and effects on the neutron form factors, especially for
polarization measurements, appear to be negligible.  However, this conclusion
is based on the assumption that these TPE corrections fully resolve the
difference.  While some calculations indicate that TPE corrections are
sufficient, direct experimental evidence for TPE effects of the necessary
magnitude is very limited.

Two-photon exchange corrections have the opposite sign in positron and
electron scattering.  While early comparisons of positron and electron
scattering generally showed very small effects, a reexamination of these data
focusing on the angular dependence of the correction found some evidence for
TPE corrections of the kind needed to resolve the
discrepancy~\cite{arrington04b}. There are currently several experiments
whose goal is to study TPE corrections, either through direct comparisons of
positron and electron scattering, or through detailed examination of the
angular dependence of cross section or polarization observables.  Other
experiments aim to extract polarization observables that are forbidden in
the Born approximation.  These observables relate to the imaginary part of the
TPE amplitude, while the form factor extractions are affected by the real
part, but these measurements will provide a clean way to test calculations of
the TPE effects.   A detailed review of our current understanding and the
ongoing experimental program designed to isolate these contributions can be
found in Ref.~\cite{carlson07,arrington07a}.

\section{Impact of the New Data}

The new high $Q^2$ data for the proton, with the dramatically different
$Q^2$ dependence for $\gep$ and $\gmp$, led to a significant reevaluation
of the high $Q^2$ region.  It was realized that while the $Q^2$ dependence
of $\gep$ was not consistent with the leading term of the pQCD prediction, the
inclusion of orbital angular momentum introduced an additional falloff which
may explain the high $Q^2$ data~\cite{belitsky03,ralston04}.

In addition, the difference in the electric and magnetic form factors implies
a difference in the spatial charge and magnetization distributions.  In the
textbook picture, the spatial distribution is related to the Fourier transform
of the form factors.  However, these spatial distributions are in the Breight
frame, and so each $Q^2$ value requires a different boost to get to
the rest frame, thus requiring model dependent relativistic boost corrections
which become large at high $Q^2$.  A comparison of the difference between
charge and magnetization distributions, applying one model of the boost
corrections, showed that the central magnetic density was roughly 50\% larger
than the electric density~\cite{kelly02}.  Other works have looked beyond
the overall density, and found non-spherical, highly complex structures in
the spatial distribution of the charge when examining quarks of specific
moment, or specific spin~\cite{belitsky04,miller03}.  These provide
additional information on the correlation of the quark space, spin, and
momentum degrees of freedom, but require a model of the full generalized
parton distribution, rather than simply the form factors.

More recently, a model-independent method was found to extract the transverse
charge distribution in the infinite momentum frame (IMF)~\cite{miller07}, the same
frame in which the longitudinal momentum distributions of the quarks can be
extracted from the parton distribution functions.  In the IMF, the transverse
charge density at the center of the nucleon is dominated by the high-$x$ pdfs,
i.e. the high momentum quarks in the nucleon, corresponding to a positive core
for the proton and negative core for the neutron.  This differs from the rest
frame distributions, where the neutron has a positive core.  Further details
of the interpretation are available for the proton~\cite{miller08a} and
neutron~\cite{miller08b}.

The neutron data were limited to a lower $Q^2$ region, but have been of equal
importance in improving our understanding of the nucleon structure. 
Calculations of the nucleon structure have historically been able to do
a fair job of reproducing $\gmp$, which was well known over a large $Q^2$
range, but the data on the other form factors has been of much lower quality,
and thus not been able to really challenge the assumptions of the models in a
global and systematic fashion.  Calculations must now explain both $\gmp$ and
the dramatically improved data on $\gep$ and $\gmn$.  The neutron electric form
factor is a special case, as one would have $\gen=0$ if the up and down quark
spatial distributions were identical, and thus it is sensitive to difference
in the flavor distributions.  Thus, the ``pion cloud'' contributions, coming
from fluctuations of the neutron into a virtual proton and negative pion, yield
a leading contribution to $\gen$.  Such fluctuations are small perturbations
to the distributions for the other form factors, although an intriguing
analysis~\cite{friedrich03} suggested that there were common features in all
four form factors that may be attributable to the pion cloud contributions.
Nonetheless, $\gen$ provides a unique sensitivity to pion contributions, which
are difficult to include in constituent quark models of the nucleon, and it is
therefore difficult to simultaneously describe all of the form factors within
a single model.

Interest in low $Q^2$ form factors has also led to efforts to make higher
precision measurements of the proton form factors at low
$Q^2$~\cite{crawford07,ron07}.  The data, along with recent experiments
that are still being analyzed, will improve the precision in the region
where the pion cloud contributions are believed to be the largest.  In
addition, high precision measurements, utilizing these new polarization
techniques and applying new corrections for two-photon exchange terms, have
an impact on other experiments.  Calculations of the hydrogen hyperfine
splitting, measured to 12 decimal places, are limited by hadronic corrections
relating to the proton size, and thus the low $Q^2$ form
factors~\cite{carlson08}.  Precise knowledge of these low $Q^2$ form factors
is also important in extracting the contribution of strange quarks to the form
factors.  The nucleon form factors are the sum of up, down, and strange
quark contributions, and by measuring the proton and neutron form factors,
along with a third set of form factors, we can extract the up, down, and
strange quark contributions.  That third set is the form factor associated
with Z-boson exchange, which can be measured in parity violating electron
scattering~\cite{beise05,young06,liu07}.  In the case of no strangeness, the
parity violating asymmetry can be calculated from just the proton and neutron
form factors, and thus provides the baseline for extracting strange quark
contributions.  As with the form factor program, this has also relied on the
development of high polarization, high intensity electron beams, and has made
dramatic progress in the last ten years.

\section{Future Plans}

The energy upgrade at Jefferson Lab will allow for dramatic extensions
of these form factor measurements to higher $Q^2$, and providing a complete
set of high-precision measurements up to relatively high $Q^2$.  In addition,
the are new measurements pushing the level of precision at low $Q^2$,
which provide useful information on the proton structure and complement
the low $Q^2$ parity violating elastic scattering measurements.  Rapid
progress is still being made in modeling form factors, and in QCD-based
calculations of the nucleon form factors.  The data acquired over the past
decade led to a dramatic resurgence in these efforts, and the promise of
future data, and the ability to apply these models to a range of form factors
and related observables, have made this an exciting and rapidly progressing
field.

\section*{Acknowledgments}

This work is supported by the U.S. Department of Energy, Office of Nuclear
Physics, under contract DE-AC02-06CH11357

\bibliography{panic08_arrington}

\end{document}